\documentclass[sigconf]{acmart}

\usepackage{booktabs} 

\setcopyright{rightsretained}

\newcommand{\s}{\mathbf{s}}

\newcommand{\x}{\mathbf{x}}
\newcommand{\h}{\mathbf{h}}
\newcommand{\Gb}{\mathbf{G}}
\newcommand{\Q}{\mathbf{Q}}

\acmDOI{}

\acmISBN{}

\acmConference[CMDWM'17]{4th Workshop on Complex Methods for Data and Web Mining}{August 2017}{Leipzig, Germany} 
\acmYear{2017}
\copyrightyear{2017}

\acmPrice{}

\begin{document}
\copyrightyear{2017} 
\acmYear{2017} 
\setcopyright{acmlicensed}
\acmConference{WI '17}{August 23-26, 2017}{Leipzig, Germany}\acmPrice{15.00}\acmDOI{10.1145/3106426.3109424}
\acmISBN{978-1-4503-4951-2/17/08}

\title{Seasonality in Dynamic Stochastic Block Models}

\author{Jace Robinson}
\affiliation{%
  \institution{Dept. of Computer Science \& Engineering \\ Kno.e.sis Research Center \\Wright State University}
  \city{Dayton} 
  \state{Ohio} 
}
\email{robinson.329@wright.edu}

\author{Derek Doran}
\affiliation{%
  \institution{Dept. of Computer Science \& Engineering \\ Kno.e.sis Research Center \\Wright State University}
  \city{Dayton} 
  \state{Ohio} 
}
\email{derek.doran@wright.edu}

\begin{abstract}
Sociotechnological and geospatial processes exhibit time varying structure that make 
insight discovery challenging. This paper proposes a new statistical 
model for such systems, modeled as dynamic networks, to address this challenge. It assumes that 
vertices fall into one of $k$ types and that the probability of edge formation at a particular
time depends on the types of the incident nodes and the current time. The time
dependencies are driven by unique seasonal processes, which many systems
exhibit (e.g., predictable spikes in geospatial or web traffic each day). The paper 
defines the model as a generative process and an inference procedure to recover 
the seasonal processes from data when they are unknown. Evaluation with synthetic 
dynamic networks show the recovery of the latent seasonal processes that drive its
formation.
\end{abstract}

%
%


\keywords{Dynamic Networks, Kalman Filter, Structural Time Series, State Space Model}

\maketitle

\section{Introduction \& Motivation}

\begin{figure}
\centering
\includegraphics[width=.5\textwidth]{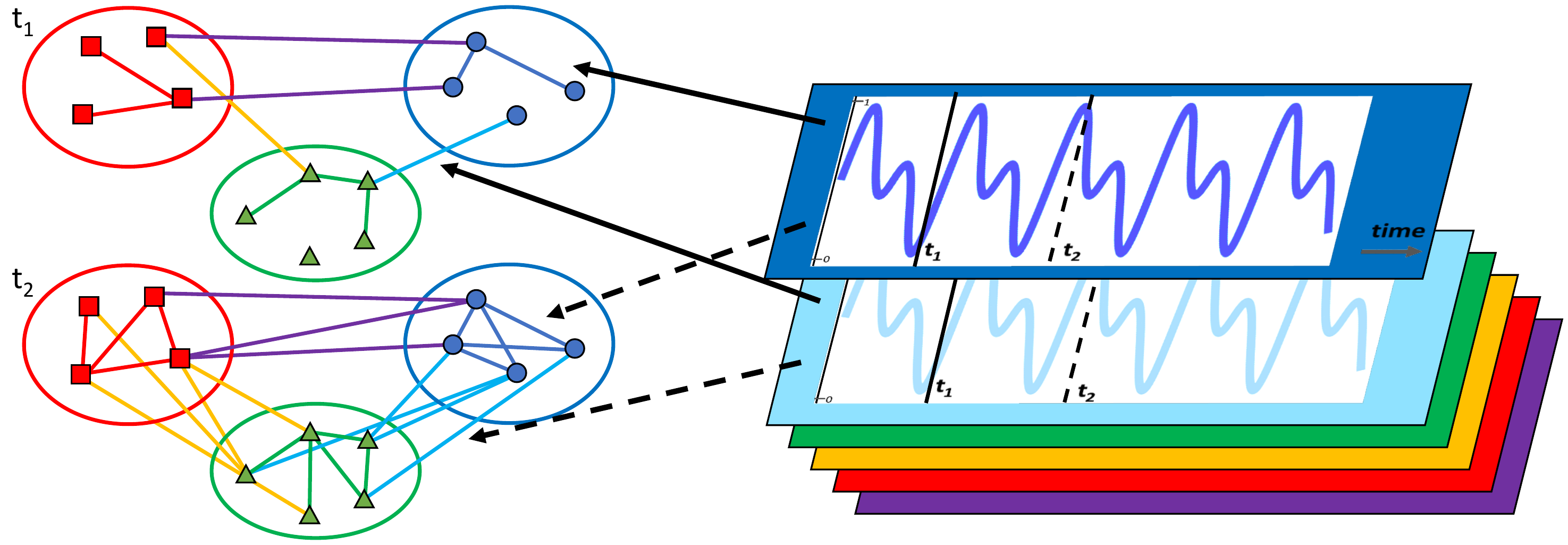}
\caption{Example of seasonality in dynamic networks. In this three class network, 
a latent seasonal process determines a probability of edge formation at times $t_1$ and $t_2$, subject
to both process and measurement noise. A different process (colored plates on the right) 
affect probabilities for edges connecting unique pairs of node types.
The paper presents a statistical model codifying these ideas, which may be useful in 
comparison, prediction, and anomaly detection tasks on dynamic complex systems.}
\label{fig:SDSBM-Overview}
\vspace{-15px}
\end{figure}

Many complex systems exhibit regular, time dependent, {\em seasonal} patterns. For example, human movement 
patterns are driven by the time of day~\cite{song2010limits}, and vehicle traffic densities exhibit 
predictable increases at certain hours causing rush hours and decreases at night~\cite{li2009temporal}. This same `seasonal', time dependent effect occurs when monitoring network bandwidth usage \cite{yoo2016time} or when counting the number of clicks per day on a web page\cite{gonccalves2008human}.

We look to bring the notion of seasonality to statistical network modeling with a new kind of dynamic stochastic block model (DSBM). A DSBM asserts
that system components (nodes) are grouped into several types, and the probability of 
observing a component relation or interaction (edges) are determined by 
the types of the incident nodes and time. Different kinds of DSBM 
consider different assumptions about the network formation 
process~\cite{xu2013dynamic, xu2015stochastic, Yang2011, xing2010state, ho2011evolving}, 
but none consider seasonality.  
A conceptual overview of the model we propose is given in Figure~\ref{fig:SDSBM-Overview}.
It fuses structural time series (plates on the right of Figure~\ref{fig:SDSBM-Overview}) with a generative network model.
We call this a seasonal DSBM (SDSBM). The model is transformed to a state-space
model for scalable fitting to data by Kalman Filters, and expectation-maximization is used for parameter learning.

\section{Model Specification} \label{sec:modelSpec}
We first specify the model of the seasonal processes controlling edge dynamics. 
We assume that time is discrete, with the current time $t$ representing a 
time period of some resolution. We also assume the node types are provided.
For each pair of node types $a$ and $b$, we consider a structural
time series with a {\em bias} $m_t^{(a,b)}$ establishing an
anchor for values of the time series and a {\em seasonal offset} $s_t^{(a,b)}$ that 
shifts the bias by the current seasonality position. The process at time $t$, denoted $c^{(a,b)}_t$, is  
\begin{equation} \label{eq:genLevel}
c^{(a,b)}_t = m^{(a,b)}_t + s^{(a,b)}_t \end{equation}
with bias $m^{(a,b)}_t$ described by
$m^{(a,b)}_t = m^{(a,b)}_{t-1} + \delta_{m^{(a,b)}_t}$ 
where $\delta_{m^{(a,b)}_t} \sim \mathcal{N}(0,q^{(a,b)}_m)$ models possible process noise. 
The set of seasonal offsets are stored in a vector
 $\s^{(a,b)} =(s^{(a,b)}_1, s^{(a,b)}_2, ..., s^{(a,b)}_d)$ having $d$ components. 
 $d$ reflects either the length or the resolution
of a seasonal process (e.g., $d=60$ to model per minute changes over a process that
cycles per hour) and is assumed to be provided by the user of the model. The components are: 
\begin{equation}\label{eq:genSeasonality}
s^{(a,b)}_t = -\sum_{i=1}^{d-1}s^{(a,b)}_{t-i} + \delta_{s^{(a,b)}_t} 
\end{equation}
where $\delta_{s^{(a,b)}_t} \sim \mathcal{N}(0,q^{(a,b)}_s)$. 
This form 
enforces a  zero-sum constraint to increase identifiability~\cite{murphy2012machine}. 
It should be emphasized that
$q^{(a,b)}_m$ and $q^{(a,b)}_s$ control the noise of the underlying seasonal process, where low noise represents a stable seasonality that does not vary significantly between realized periods.

To model how the seasonal processes govern the shape of a dynamic network, 
we define a random variable $E^{(a,b)}_t \in [0,1]$ as the {\em expected density} of edges spanning node types $(a,b)$ at time $t$:
\begin{equation}\label{eq:genDensity}
E^{(a,b)}_t = c_t^{(a,b)} +  \epsilon_{E^{(a,b)}_t}
\end{equation}
where $\epsilon_{E^{(a,b)}_t} \sim \mathcal{N}(0,r^{(a,b)})$ models possible
measurement noise. 
Next we define an adjacency matrix $A_t$, where $[A_t]_{ij} = 1$ if there exists an edge between nodes $i$ and $j$ at time $t$ and $[A_t]_{ij}=0$ otherwise. 
Denote $A_t^{(a,b)}$ as the submatrix of $A_t$ only containing the rows and columns representing type $a$ and type $b$ nodes. Then $A_t^{(a,b)}$ is defined by the random variable:
\begin{equation} \label{eq:genAdj}
[A_t^{(a,b)}]_{ij} \sim Bernoulli(E_t^{(a,b)})
\end{equation}
Repeating this process for all blocks $(a,b)$ and time steps $t$ will generate a desired dynamic network $\mathcal{A} = \{A_1, A_2, ..., A_t\}$.

\section{Model Fitting}
We now describe an inference procedure to fit the model to an observed $\mathcal{A}$. 
Since seasonal processes are latent in data, the task is to estimate a posterior distribution 
on each $m^{(a,b)}_t$ and $s^{(a,b)}_t$ for each pair of node types $(a,b)$. 
Kalman filters~\cite{kalman1960new} are an appropriate tool for this task, but requires 
transforming the generative model into a state-space model (SSM). A SSM 
is a time series model with \textit{hidden} and \textit{observed} variables~\cite{murphy2012machine}; 
here we define $\x_t$ as hidden and $w_t$ as observed variables, respectively. We can notice the bias and seasonal offsets from before were hidden, while the adjacency matrix is observed, foreshadowing the structure to be defined. 
A SSM creates observations at time $t$ by two linear models: 
An {\bf observation model}  $w_t = \h \x_{t} + \epsilon_t$ and 
a {\bf transition model} $\x_t = \Gb \x_{t-1} + \Delta_t$.
Here, observations $w_t$ are generated by a transformation of the output 
(defined by $\h$) of the underlying transition model. The transition model describes transformations within a hidden state space where transitions 
from time $t-1$ to time $t$ are defined by the matrix $\Gb$. Observations and transitions are
to be subject to time dependent random noise, which are modeled by Gaussian distributions
$\epsilon_t \sim \mathcal{N}(0, R_t)$ and $\Delta_t \sim \mathcal{N}(0, \Q_t)$
With $R_t$ and $\Q_t$ controlling the amount of observation and transition noise, respectively.
Assuming parameters $\theta_t = \{\h, \Gb, R_t, \Q_t\}$ are known, a Kalman Filter can be used to derive the exact posterior $Pr(\x_t | \theta_t, w_1, w_2, ...,w_t)$, i.e., the probability of the hidden state value at time $t$ given the observations up to and including time $t$~\cite{murphy2012machine}. 
 
Now we transform the model specification into a state space to define the transition 
model $\x_t = \Gb \x_{t-1} + \Delta_t$. As we are assuming edges of different vertex types $(a,b)$ are independent of each other, we will formulate the inference in terms of a pairing $(a,b)$. The full inference is completed by repeating the process for all pairs $(a,b)$.  
First we transform Equations~\ref{eq:genLevel} and~\ref{eq:genSeasonality} to define the 
hidden state variable $\x_t$ and state transition $\Gb$. The hidden state will be composed of the bias and vector of seasonal offsets as a $d \times 1$ seasonal state vector for a 
period of length $d$: 
\begin{equation}
\x^{(a,b)}_t = \begin{bmatrix}
    m^{(a,b)}_t&
    s^{(a,b)}_t&
    s^{(a,b)}_{t-1}&
    \hdots&
    s^{(a,b)}_{t-d+2}
\end{bmatrix}^T
\end{equation}
Note that all the seasonal offsets from $\s^{(a,b)}$ are maintained in the state for a given $t$, with the $d$th seasonal offset implicitly defined based on the zero-sum constraint. Now to perform the state transition from time $t-1$ to time $t$ we define a $d\times d$ matrix $\Gb$:

\begin{equation} \label{eq:gb}
\Gb =  \begin{bmatrix}
    1 & 0 & 0 & \dots & 0 & 0 \\
    0 & -1 & -1 & \dots & -1 & -1 \\
    0 & 1 & 0 & \dots & 0 & 0 \\
    0 & 0 & 1 & \dots & 0 & 0 \\
    \vdots & \vdots & \vdots & \ddots & \vdots & \vdots \\
    0 & 0 & 0 & \dots & 1 & 0
\end{bmatrix}
\end{equation}
In $\Gb$, we see that multiplication of the first row of $\Gb$ by $\x^{(a,b)}_t$ yields Equation~\ref{eq:genLevel} without noise being added, as only the bias term is updated.
Multiplication of the second row of $\Gb$ by $\x^{(a,b)}_t$ will update a \textit{single} seasonal offset as shown in Equation~\ref{eq:genSeasonality}. The remaining rows of $\Gb$ serve to permute the remaining seasonal offsets, such that each offset $s^{(a,b)}_i$ is updated after a full period of $d$ time steps. Each time step will update the most current seasonal offset, and shift the remaining offsets right one index in the state vector. Next we define the $d \times 1$ noise vector $\Delta_t = [ \delta_{m^{(a,b)}_t}, \delta_{s^{(a,b)}_t}, 0, \dots, 0 ]^T$ where the first element is the bias noise in Equation~\ref{eq:genLevel}, the second element is seasonal noise in Equation~\ref{eq:genSeasonality}, 
and the remaining elements are all 0 as there is no additional noise for permuting the seasonal offsets.  These noise values are sampled from a zero mean Gaussian with $d \times d$ covariance matrix $\Q = \texttt{diag}[q_m^{(a,b)}, q_s^{(a,b)},0,\dots,0]$. Assuming the bias noise $\delta_{m^{(a,b)}_t}$ and seasonal offset noise $\delta_{s^{(a,b)}_t}$ are independent, the off-diagonal elements of $\Q$ are zero. The remaining elements are all zero, reflecting the lack of noise for the permutation operations. We can assume $\Q$ is stationary, and drop the dependence on $t$. This complete formulation of the transition model is not new, and has been completed by other researchers such as in \cite{davis2005modeling}. 


Our next task is to transform Equations~\ref{eq:genDensity} and~\ref{eq:genAdj} into the
observation model $w_t = \h \x_{t} + \epsilon_t$. To do this, we will need to define some additional 
variables, and take advantage of a result of the central limit theorem for a large number of vertices with types $(a,b)$
to create an approximate Gaussian transformation. First we need a count of the number of \textit{possible edges} in block $(a,b)$, so if there are $|a|$ nodes of type $a$ and $|b|$ nodes of type $b$ then define:
\begin{equation}\label{eq:n}
   n^{(a,b)}=\begin{cases}
   a=b & \frac{|a|(|a|-1)}{2} \\
   a\neq b &  |a||b|\\
   \end{cases}
\end{equation}
Also define the random variable $p_t^{(a,b)} \sim Binomial(E^{(a,b)}_t, n^{(a,b)})$ as the number 
of {\em formed edges} 
in block $(a,b)$ at time $t$, where $E^{(a,b)}_t$ is the expected edge density as 
determined by Equation~\ref{eq:genDensity}. $p_t^{(a,b)}$ is simply a more mathematically convenient way to define the edge generation process from Equation~\ref{eq:genAdj} and does not change the overall model. Having a binomial random variable, for large enough $n^{(a,b)}E_t^{(a,b)}$ we can apply 
the central limit theorem to approximate the distribution of $p_t^{(a,b)}$ as Gaussian:
 \begin{equation} \label{eq:sumEdges}
 	p_t^{(a,b)} = n^{(a,b)}E_t^{(a,b)} +  \omega_{p^{(a,b)}_t}
 \end{equation}
where $\omega_{p^{(a,b)}_t} \sim \mathcal{N}(0, n^{(a,b)}E_t^{(a,b)}(1-E_t^{(a,b)}))$  is {\em observation noise} that is time dependent on $E_t^{(a,b)}$. This represents variation in the repeated 
binary decision process of whether pairs of complex system components will interact. For example, in a geospatial context, a seasonal process may dictate that people travel from home to work at 8am, 
yet people individually decide the precise time they leave for work. In this example, locations of the geospace are system components (vertices), and movement between these locations are interactions (edges). In the SDSBM, each system component will have a type, such that each individual home is a vertex and all these homes can have the same type of `residence'. The sum of individual 
departure time decisions creates variation captured in $\omega_{p^{(a,b)}_t}$. 

Now that we have successfully transformed our edge sampling procedure from Equation~\ref{eq:genAdj} to an approximately Gaussian formulation in Equation~\ref{eq:sumEdges}, 
we can return to defining observation model parameters $w_t$ and $\h$. We will define the observed variable $w^{(a,b)}_t$ as the number of formed edges  $p_t^{(a,b)}$. To create $\h$, 
define a transformation which takes as input a seasonal state vector $\x_t^{(a,b)}$ and produces as output the number of formed edges $w^{(a,b)}_t$. By combining the operations of Equations~\ref{eq:genDensity} and~\ref{eq:sumEdges} we define:
\begin{equation}
\h = \begin{bmatrix}
    n^{(a,b)} & n^{(a,b)} & 0 & \dots & 0
\end{bmatrix}
\end{equation}
Examining $w^{(a,b)}_t = \h\x^{(a,b)}_t$ closer, we can see this multiplication both sums the bias $m_t^{(a,b)}$ and first seasonal offset $s^{(a,b)}_t$ following Equation~\ref{eq:genDensity}, 
and multiplies by $n^{(a,b)}$ to match the repeated Bernoulli trials from 
Equation~\ref{eq:sumEdges}. Finally, we model the variance of the 
measurement noise $\epsilon_t$ by summing measurement noise
$\epsilon_{E_t^{(a,b)}}$ and observation noise $\omega_{p^{(a,b)}_t}$. This definition 
allowing for more flexible modeling of many complex systems. $\epsilon_t$ is sampled from a zero mean Gaussian distribution with time dependent variance $R^{(a,b)}_t = n^{(a,b)}E_t^{(a,b)}(1-E_t^{(a,b)}) + (n^{(a,b)})^2r^{(a,b)}$.

We have now transformed the generative procedure to the suitable SSM to allow easy inference via the Kalman Filter. Given an initial Gaussian belief state $Pr(\x^{(a,b)}_0)$ with mean $\x_0^{(a,b)}$ and variance $\Sigma_0^{(a,b)}$, all subsequent belief states will be Gaussian as well. The closed-form updates for the posterior distribution \\$Pr(\x^{(a,b)}_t | \theta^{(a,b)}_t, w^{(a,b)}_1, w^{(a,b)}_2, ...,w^{(a,b)}_t)$ are not defined in this paper due to space constraints but
are available in~\cite{murphy2012machine}. 

To estimate the unknown noise parameters of the SSM and Kalman Filter $\theta_t = \{R_t, \Q\}$ from data, we derive an 
expectation-maximization routine that iteratively converges to locally optimal point estimates. The update equations for $\Q$ can be found in other research such as \cite{harvey1990estimation}. The updates for $R_t$ is a new formulation, which deviates from the conventional EM routine for Kalman Filters, due to our separation of noise parameter $R^{(a,b)}_t$ into a combination of $E_t^{(a,b)}$, $n^{(a,b)}$, and $r^{(a,b)}$. $n^{(a,b)}$ is assumed fixed and $E_t^{(a,b)}$ can be estimated using the prediction step of the Kalman Filter as described in \cite{xu2013dynamic}. To estimate $r^{(a,b)}$, we set up an optimization routine, which maximizes the log-likelihood of the complete joint distribution $Pr(\x^{(a,b)}_1, ..., \x^{(a,b)}_t, w^{(a,b)}_1,...,w^{(a,b)}_t)$. This optimization is completed each iteration of EM, until a locally optimal estimate for both $\Q$ and $R_t$ is found.

\section{Results}
\begin{figure}
\centering
\includegraphics[width=.5\textwidth]{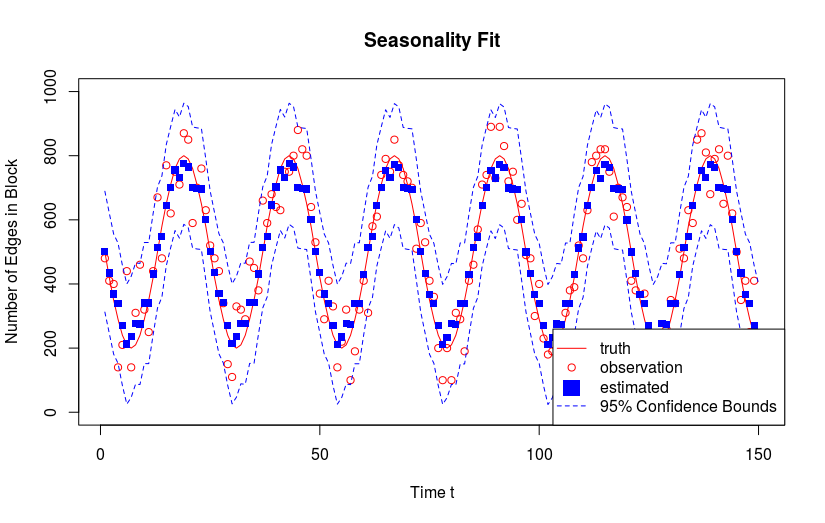}
\vspace{-25px}
\caption{Comparison of the inferred seasonality (posterior distribution) compared to actual seasonality governing connections between nodes of type $x$ and $y$. The underlying seasonality follows a simple sine wave that is accurately recovered from noisy observations.}
\vspace{-5px}
\label{fig:sdsbm-example}
\end{figure}

\begin{figure}
\centering
\includegraphics[width=.5\textwidth]{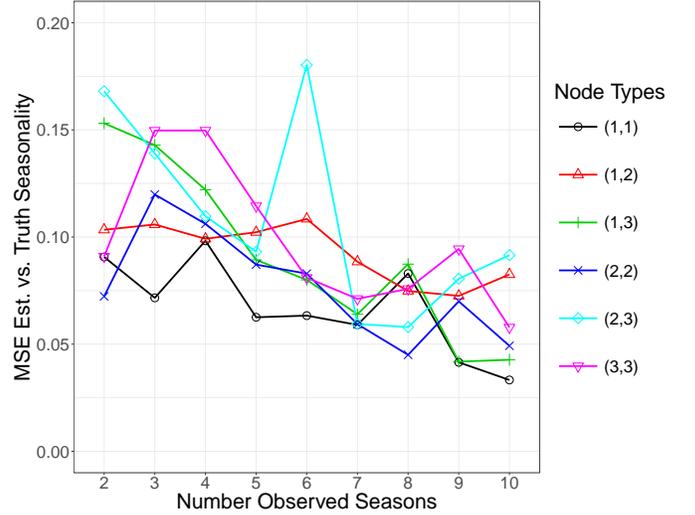}
\caption{The MSE between the estimated seasonal state vectors to the true seasonal state vectors as a function of the number of seasonal 
periods observed. The error decreases linearly with observation length.}
\label{fig:obsSeasons}
\vspace{-10px}
\end{figure}

\begin{figure}
\centering
\includegraphics[width=.45\textwidth]{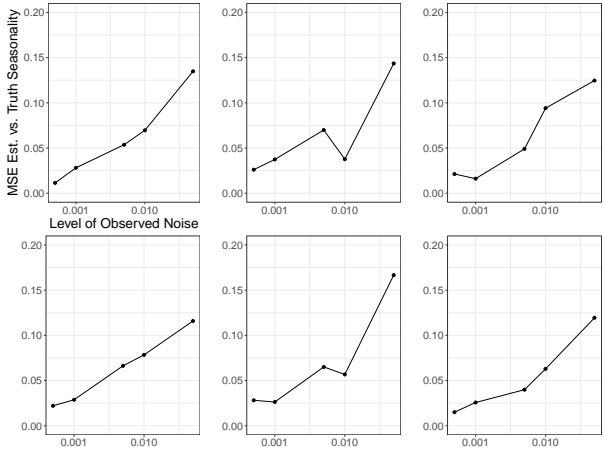}
\caption{The MSE between the estimated seasonal state vectors to true seasonal state vectors as we increase the amount of observation noise $r_t$
for each pair of node types. The x-axis is log scale while the y-axis is continuous. The errors appear to increase exponentially with the level of observation noise.}
\label{fig:obsNoise}
\vspace{-20px}
\end{figure}

To demonstrate the fidelity of the SDSBM, we perform several experiments on a synthetic dynamic network  $\mathcal{A} = \{A_1, A_2, ..., A_T\}$ generated following the process defined in Section~\ref{sec:modelSpec}. The network is defined with $k=3$ vertex types, creating $6$ unique pairs $(a,b)$. There are approximately $32$ nodes of each type, creating the number of possibles edges for each $(a,b)$ as $n^{(a,b)}=1000$. All pairings $(a,b)$ have the same underlying true seasonality. 
We start with an initial bias $m^{(a,b)}_0$ and initial vector of seasonal offsets $\s^{(a,b)} = (s_1^{(a,b)}, s_2^{(a,b)}, ..., s_d^{(a,b)})$. We set $m_0^{(a,b)} = 0.5$, and the seasonal offsets as samples from a sine wave with amplitude $0.3$. This will create expected block densities $E_t^{(a,b)} \in [.2, .8]$, where noisy samples may cover the full range $[0,1]$. For example, in some of the experiments, we set the length of seasonality as $8$, creating $\s^{(a,b)} = (.2,.3,.2,0, -.2, -.3, -.2,0)$. To determine the noise level of the system, we set bias variance $q^{(a,b)}_m= 10^{-8}$ and seasonal offset variance $q^{(a,b)}_s = 10^{-8}$. These are set very low to reflect the expectation of near constant seasonality. If we set the number of possible edges of types $(a,b)$ as $n^{(a,b)} = 1000$, we define a `medium' amount of noise at $r^{(a,b)} = 5.5 * 10^{-3}$, which will result in a standard deviation of the number of formed edges of $75$. Repeating the operations in Equations~\ref{eq:genLevel}-\ref{eq:genAdj} $T$ times will form the desired synthetic dynamic network $\mathcal{A}$.

To infer the underlying seasonal processes, thus fitting the model to $\mathcal{A}$, the user needs to provide values for seven hyperparameters. We assume the user knows both the node labels and desired length of seasonality $d$. Initial guesses need to be provided for the remaining hyperparameters of initial state mean $\x_0^{(a,b)}$, state variance $\Sigma_0^{(a,b)}$, measurement variance $r^{(a,b)}$, bias variance $q_m^{(a,b)}$, and seasonal offset variance $q_s^{(a,b)}$. These can be defaulted to initial values such as $1$, without having a significant consequence on the fidelity performance. A more formal exploration of the sensitivity of the model to these hyperparameters is left as future work.
We present the results of three experiments. First we simply infer the hidden seasonal state vectors $\x_t^{(x,y)}$ for a specific pairing $(x,y)$ and for all $t$, to qualitatively demonstrate the goodness of fit of the algorithm. A visual of the seasonal state vector transformed to expected edge counts is given in 
Figure~\ref{fig:sdsbm-example}. The underlying seasonality is estimated well, as the estimates shown as blue squares, closely follow the true seasonality of the solid red line. 
In the second experiment, we look to demonstrate the model's ability to recover the seasonal state vector with differing numbers of observations as controlled by $T$, the length of the dynamic network. We start with $T=2d$ and increase the number of observed periods until $T=10d$. To assess performance we calculated the mean-squared error (MSE) between the true $\x_t^{(a,b)}$ and estimated $\hat{\x}_t^{(a,b)}$ seasonal vectors for all $t$. In Figure~\ref{fig:obsSeasons} a negative linear dependence between MSE and number of observations is demonstrated in all pairs $(a,b)$. The model becomes increasingly accurate for larger datasets. 
In the third experiment, we again evaluate the MSE between truth and estimated seasonality, but this time we vary the amount of measurement noise $r^{(a,b)}$. We start with a low $r^{(a,b)} = 5*10^{-4}$ and increasing a half order of magnitude (e.g., $5*10^{-4}$,$1*10^{-3}$, $5*10^{-3}$, etc...) until a high level of noise at $5*10^{-2}$. In Figure~\ref{fig:obsNoise} we see the exponential dependence on the noise level. The performance of this model on real data may thus 
be sensitive to this measurement noise value.

\section{Conclusion}
This paper proposed a new statistical model for dynamic networks, leveraging the benefits of structural time series and stochastic block models. The generative specification and inference procedure are defined. We demonstrate the capabilities of the model on a synthetic dataset, showing some properties of the model such as the negative linear dependence to the number of observed seasonal periods.

\begin{acks}
This work is supported by industry and government partners at the National Science Foundation's I/UCRC Center for Surveillance Research and the Air Force Research Laboratory.
\end{acks}

\bibliographystyle{ACM-Reference-Format}
\bibliography{CMDWM-ref} 

\end{document}